\pdfoutput=1
\documentclass[%
 reprint,
superscriptaddress,
 amsmath,amssymb,
 aps,
]{revtex4-2}
\usepackage[utf8]{inputenc}
\usepackage[T1]{fontenc}
\usepackage[english]{babel}
\usepackage{verbatim}
\usepackage{microtype} 

\usepackage{xcolor} 
\usepackage{lipsum} 
\usepackage{dcolumn}
\usepackage{bm}
\usepackage{graphicx} 
    \graphicspath{ {./figures/} } 
\usepackage{textcomp} 
    \usepackage{gensymb} 
\usepackage{siunitx} 
\usepackage{csquotes} 

\usepackage{hyperref} 



\begin{document}

\preprint{APS/123-QED}

\title{Cross-architecture Tuning of Silicon and SiGe-based Quantum Devices \protect\\ Using Machine Learning}

\author{B. Severin}
    \affiliation{Department of Materials, University of Oxford, Parks Road, Oxford, OX1 3PH, UK}
    
\author{D. T. Lennon}
    \affiliation{Department of Materials, University of Oxford, Parks Road, Oxford, OX1 3PH, UK}
\author{L. C. Camenzind}
    \affiliation{Department of Physics, University of Basel, Basel, 4056, Switzerland}%
\author{F. Vigneau}
    \affiliation{Department of Materials, University of Oxford, Parks Road, Oxford, OX1 3PH, UK} 
\author{F. Fedele}
    \affiliation{Department of Materials, University of Oxford, Parks Road, Oxford, OX1 3PH, UK} 

\author{D. Jirovec}
    \affiliation{
 Institute  of  Science  and  Technology  Austria,  Am  Campus  1,  3400  Klosterneuburg,  Austria
}
\author{A. Ballabio}
    \affiliation{
 L-NESS, Dipartimento di Fisica, Politecnico di Milano, Polo di Como, ViaAnzani 42, 22100 Como, Italy
}
\author{D. Chrastina}
    \affiliation{
 L-NESS, Dipartimento di Fisica, Politecnico di Milano, Polo di Como, ViaAnzani 42, 22100 Como, Italy
}
\author{G. Isella}
    \affiliation{
 L-NESS, Dipartimento di Fisica, Politecnico di Milano, Polo di Como, ViaAnzani 42, 22100 Como, Italy
}

\author{M. de Kruijf}
    \affiliation{Department of Physics, University of Basel, Basel, 4056, Switzerland}%
\author{M. J. Carballido}
    \affiliation{Department of Physics, University of Basel, Basel, 4056, Switzerland}%
\author{S. Svab}
    \affiliation{Department of Physics, University of Basel, Basel, 4056, Switzerland}%
\author{A. V. Kuhlmann}
    \affiliation{Department of Physics, University of Basel, Basel, 4056, Switzerland}%
\author{F. R. Braakman}
    \affiliation{Department of Physics, University of Basel, Basel, 4056, Switzerland}%

\author{S. Geyer}
    \affiliation{Department of Physics, University of Basel, Basel, 4056, Switzerland}%
\author{F. N. M. Froning}
    \affiliation{Department of Physics, University of Basel, Basel, 4056, Switzerland}%

\author{H. Moon}
    \affiliation{Department of Materials, University of Oxford, Parks Road, Oxford, OX1 3PH, UK} 
\author{M. A. Osborne}
    \affiliation{Department of Engineering, University of Oxford, Walton Well Road, Oxford, OX2 6ED, UK}
\author{D. Sejdinovic}
    \affiliation{Department of Statistics, University of Oxford, 24-29 St Giles, Oxford, OX1 3LB, UK}
    
\author{G. Katsaros}
    \affiliation{
 Institute  of  Science  and  Technology  Austria,  Am  Campus  1,  3400  Klosterneuburg,  Austria
}%
\author{D. M. Zumbühl}
    \affiliation{Department of Physics, University of Basel, Basel, 4056, Switzerland}%
\author{G. A. D. Briggs}
    \affiliation{Department of Materials, University of Oxford, Parks Road, Oxford, OX1 3PH, UK}
\author{N. Ares}
    \affiliation{Department of Materials, University of Oxford, Parks Road, Oxford, OX1 3PH, UK}

\date{\today}

\begin{abstract}
    The potential of Si and SiGe-based devices for the scaling of quantum circuits is tainted by device variability. Each device needs to be tuned to operation conditions. We give a key step towards tackling this variability with an algorithm that, without modification, is capable of tuning a 4-gate Si FinFET, a 5-gate GeSi nanowire and a 7-gate SiGe heterostructure double quantum dot device from scratch. We achieve tuning times of 30, 10, and 92 minutes, respectively. The algorithm also provides insight into the parameter space landscape for each of these devices. These results show that overarching solutions for the tuning of quantum devices are enabled by machine learning.
    
\end{abstract}

\maketitle

\section*{Introduction}
    Before we can use a quantum computer we first need to be able to turn it on. There are many stages to this initial step, particularly for quantum computing architectures based on semiconductors. Silicon and SiGe devices can encode promising spin qubits \cite{Loss1998}, demonstrating excellent fidelities, long coherence times and a pathway to scalability \cite{Veldhorst2017, vandersypen2017interfacing, Watson2018, Hendrickx2020, Scappucci2020a, Takeda2021}. Many of these key characteristics revolve around the material itself providing the opportunity to be purified to a near-perfect magnetically clean environment resulting in very weak to no hyperfine interactions. As the material of choice of the microelectronics industry, gate-defined quantum dots in silicon and SiGe have great potential for the fabrication of circuits consisting of a large number of qubits, an essential requirement to achieving a universal fault-tolerant quantum computer~\cite{Fowler2012, Preskill2018}.

    Despite these ideal traits, material defects and fabrication inaccuracies result in discrepancies between device operating conditions. Multiple gate electrodes provide the ability to tune differing devices into similar operating regimes. These gate voltages define a big parameter space to be explored; a time-consuming process if carried out manually and certainly not scalable for circuits with millions of qubits. This tuning, which used to rely on experimentalists' intuition and knowledge of particular devices, can be automated using machine learning \cite{Ares2021}. The development of machine learning algorithms for quantum device tuning is even more challenging when looking for overarching solutions, successful on very different types of devices. 
    Of all the automatic approaches to tune quantum devices that have been demonstrated~\cite{Baart2016, Botzem2018, VanDiepen2018, Teske2019, Volk2019,  Mills2019, Kalantre2019, Zwolak2020, Durrer2020, Darulova2019}, as far as we know ours is the first that is versatile across different devices architectures and material systems.
    Here we present a machine learning-based algorithm, which we call `Cross-Architecture Tuning Solution using AI' (CATSAI), able to tune quantum dots in three different device architectures and material systems. This algorithm, based on an approach that allowed for the super-coarse tuning of double quantum dots defined in GaAs heterostructures \cite{Moon2020}, has the ability of being able to adapt the parameter space exploration to the type of device to be tuned. The origin and gate voltage sweep directions can be arbitrarily selected for devices operating with accumulation or depletion mode gate electrodes, and either holes or electrons as majority charge carriers. An advanced signal processing classification method handles charge switches and other noise patterns. 
    
    We demonstrate our CATSAI algorithm for a Si accumulation-mode ambipolar FinFET \cite{Kuhlmann2018, Camenzind2021a, Geyer2021}, a depletion-mode Ge/Si core/shell nanowire \cite{Froning2018,Froning2021a, Froning2021} and a laterally-defined device in a SiGe heterostructure, all operating with holes as charge carriers. We show that CATSAI outperforms random search and human experts on all devices. The demonstration of a general algorithm for the automatic tuning of devices compatible with industry manufacturing standards opens the path to building quantum circuits at scale for the next generation of quantum computers. 

\section*{Methods}
    
    \begin{figure*}[t]

        \includegraphics[scale=0.9]{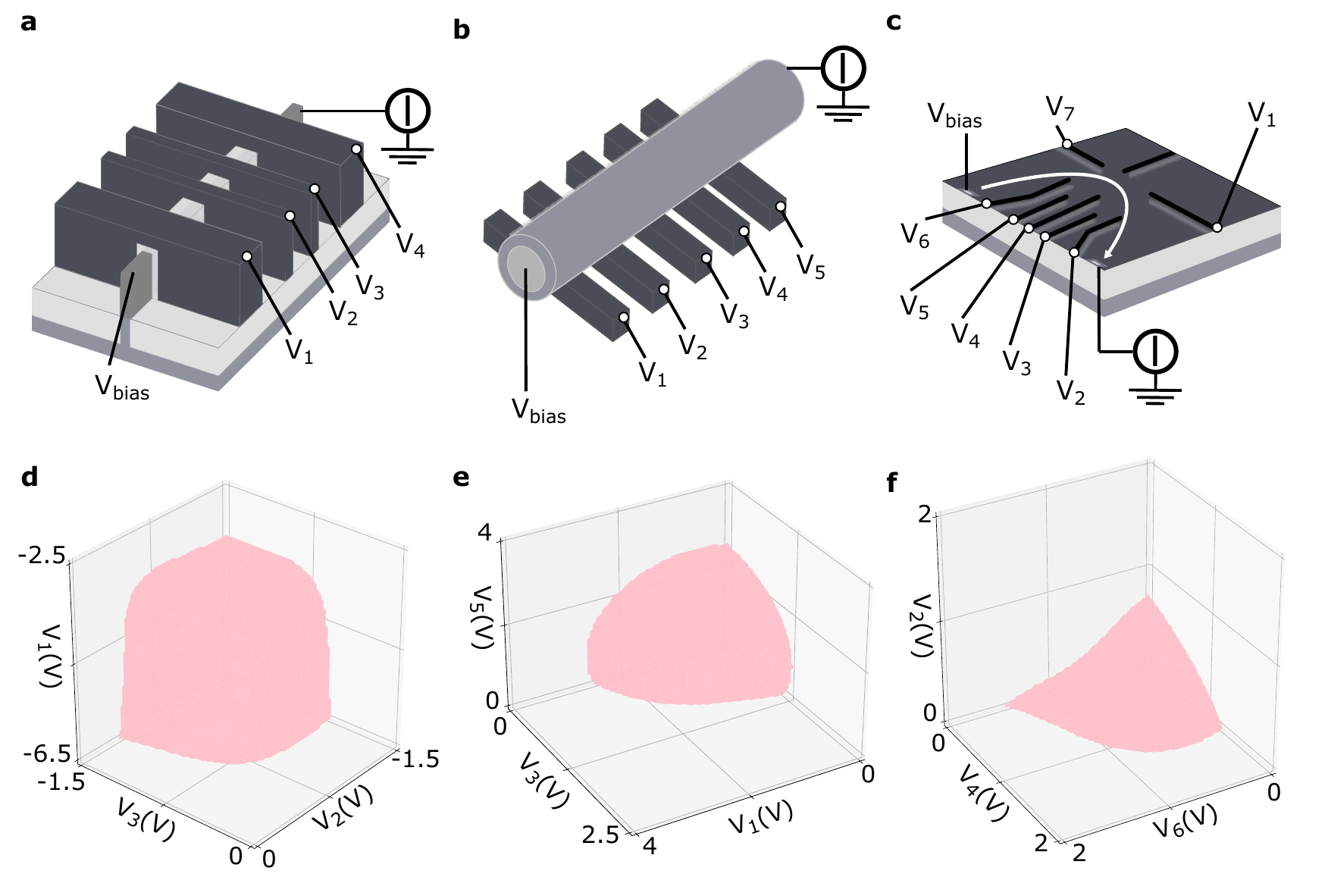}

        \caption{Device schematics. Si FinFET (a), GeSi nanowire (b) and SiGe heterostructure (c) device architectures and their corresponding current pinch-off hypersurfaces for hole transport calculated using a Gaussian process model for one of the tuning algorithm runs (d, e, f). Three gates are plotted for illustrative purposes with the remaining gates on each device set to a constant value. The bias was kept constant throughout the experiment. CATSAI was given control over the gate electrodes $V_{1}$ - $V_{4}$, $V_{1}$ - $V_{5}$, and $V_{1}$ - $V_{7}$ on the FinFET, nanowire and heterostructure, respectively.
        }
    
        \label{fig:fig1_devices}
    \end{figure*}
    \subsection*{The devices}
    Double quantum dots are defined by applying DC voltages to the gate electrodes $V_{1} - V_{4}$ for the FinFET, $V_{1} - V_{5}$ for the nanowire, $V_{1} - V_{7}$ for the heterostructure (Fig. \ref{fig:fig1_devices}). For the FinFET, the lead gate electrodes $V_{1}$ and $V_{4}$, open and close the quasi 1D silicon channel to charge carriers by controlling the size of the tunnel barrier between the quantum dots and the source and drain. The left and right plunger gate electrodes $V_{2}$ and $V_{3}$, control the occupation of the left and right quantum dot respectively. A current is driven through the FinFET by applying a bias voltage $V_{\mathrm{bias}}$ of 7.6 mV (+ 3.8 mV at the source, - 3.8 mV at the drain) to NiSi contacts \cite{Kuhlmann2018a}. The gate voltages of the FinFET are operated such that the charge carriers are holes confined by accumulation. For the nanowire, gates $V_{2}$ and $V_{4}$ act as left and right plunger gates for the quantum dots formed within the 1D channel with the remaining gates mainly controlling the tunnel barriers. Hole quantum dots are formed in depletion mode. We set $V_{\mathrm{bias}}=4$~mV. For the SiGe heterostructure, $V_{5}$ and $V_{3}$ operate as the left and right plunger gate electrodes respectively, with the remaining gate electrodes utilised as barrier gates. The white arrow denotes the flow of current. We set $V_{\mathrm{bias}}=0.5$~mV and the charge carriers are holes confined in depletion mode. The values of $V_{\mathrm{bias}}$ are set to be above typical charging energies for single quantum dots in each device. The choice of $V_{\mathrm{bias}}$ can be left to an optimiser. For the heterostructure experiments were performed at 300 mK, for the nanowire at 1.5 K and for the FinFET at 800 mK. 
    
    Voltages applied to the gate electrodes of the devices can cause the current flow to pinch-off, transitioning from a relatively high current to a near-zero value. These voltages where pinch-off occurs define a hypersurface within the entire voltage space for each device. CATSAI has no knowledge of the device architecture and generates a model of the hypersurface after a given number of iterations. The resulting hypersurface for different devices is shown in Fig.~\ref{fig:fig1_devices}d--f. Three gates are plotted for the ease of visualisation and the remaining gates are kept constant at their average value at pinch-off across the hypersurface (see Supplementary Material). 
    The hypersufaces corresponding to different devices present different curvatures, leading to different tuning landscapes. The FinFET hypersurface (Fig. \ref{fig:fig1_devices}d) is near symmetrical in the plunger gates plane, $V_{2} - V_{3}$. This is expected as these gate electrodes are nominally identical. Although $V_{1}$ is wider than the plunger gates, its effect is not stronger. The curvature of the nanowire's hypersurface is similar in the planes $V_{1} (V_{5}) - V_{3}$, since these planes are defined by the outer-middle barrier gates (Fig. \ref{fig:fig1_devices}e).
    The heterostructure's hypersurface has almost planar dependence on gate voltages $V_{2,4,6}$ (Fig.~\ref{fig:fig1_devices}f). The hypersurface's curvature in the $V_{2}$ -- $V_{4}$ plane is evidently similar to that in the $V_{6}$ -- $V_{4}$ plane, in agreement with the gate architecture. This hypersuface is qualitatively different to that reported in Ref.~\cite{Moon2020} for a relatively similar gate architecture patterned on a different heterostructure (AlGaAs/GaAs).
    The more pronounced curvature of the hypersufaces corresponding to the FinFET and the nanowire are expected given the larger gate couplings that are typically observed in these devices.
    Hypersurface characterisation could be used to inform device design and quantify device variability. 
    Despite the stark differences in gate voltage landscapes, which evidence the difficulties of cross-architecture tuning, CATSAI is able to tune across all three device architectures.
    
    \begin{figure}[t]
    \includegraphics[width=0.99\columnwidth]{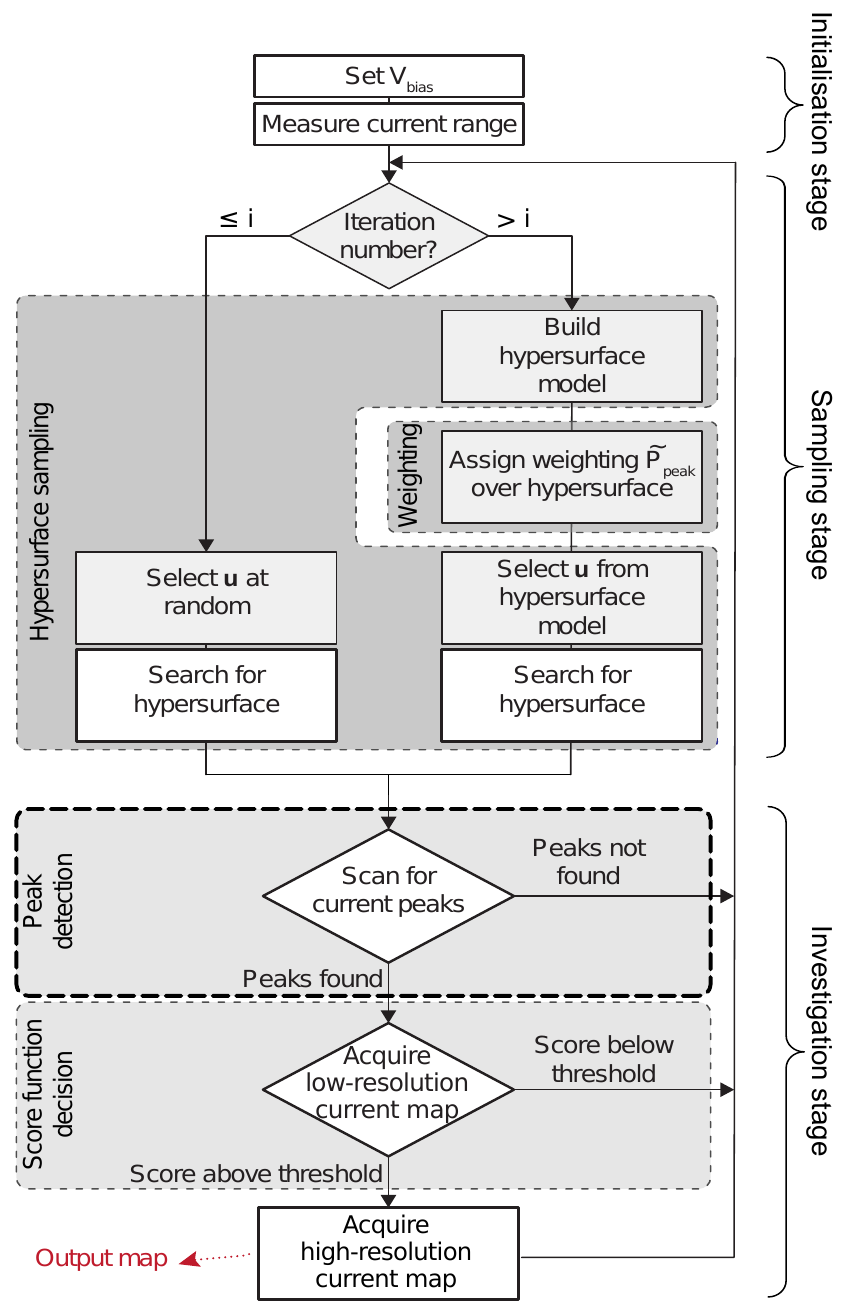}
    \caption{\label{fig:fig2_algo_workflow}
    CATSAI's workflow. The initialisation stage consists of setting $V_{\mathrm{bias}}$ then measuring the maximum and minimum current flowing through the device. The sampling stage detects pinch-off locations in gate voltage space. For the first i iterations (left-hand branch of the sampling stage), the algorithm selects $\bm{u}$ at random and travels along it until the hypersurface is found. After the i$\mathrm{^{th}}$ iteration (right-hand branch of the sampling stage), the algorithm selects $\bm{u}$ based on the model it generates of the hypersurface and of the probability of finding Coulomb peaks in a given location in gate voltage space, $\mathrm{\tilde{P}_{peaks}}$. In the investigation stage the algorithm sweeps the plunger gates to generate current traces and low-resolution and high-resolution current maps if the conditions are satisfied. The peak detection is a random forest classifier which determines whether Coulomb peaks are present or not within a current trace. After the investigation stage, the algorithm returns to the sampling stage. In each iteration, the algorithm outputs a high-resolution current map if acquired.} 
    \end{figure}
    
    \subsection*{The CATSAI algorithm}
    
    CATSAI's workflow consists of three stages, the initialisation stage, the sampling stages and the investigation stage (Fig. \ref{fig:fig2_algo_workflow}). In the initialisation stage $V_{\mathrm{bias}}$ is fixed, and the current range, i.e. the maximum and minimum current flowing through the device, is determined by measuring the current both with all the gate electrodes set to 0 V and to their maximum permissible magnitude. To avoid damage to the device the algorithm is given voltage bounds in which it can operate each gate electrode (see Supplementary Material). 
    After the initialisation stage, the algorithm turns to the sampling stage. Since the algorithm is unaware of the characteristics of the device, for the first i iterations of the sampling stage, the algorithm selects a vector $\bm{u}$ at random in the gate voltage space of the device. This vector consists of all the gate voltages considered for tuning. The algorithm then sweeps the gate voltages along that direction until pinch-off occurs. The algorithm identifies the onset of pinch-off as a current drop below a certain threshold ($0.5\%$ of the measured current range). The $N$-dimensional hypersurface is delimited by the pinch-off voltages of the $N$ gate electrodes for each device.
    
    At the start of the investigation stage, once pinch-off is found in a given gate voltage direction, a high-resolution current trace is performed. This current trace, which starts at the pinch-off location and runs diagonal to the plane defined by the plunger gates, was set to have a fixed length of 128 pixels and resolution 1.56~mV/pixel for the nanowire and 0.78~mV/pixel for the FinFET and the heteroturecture. The plunger gates, selected before running the algorithm, are those expected to predominantly shift the electrochemical potential in left and right dots.
    Using a random forest classifier \cite{Breiman2001, scikit-learn}, the algorithm determines whether Coulomb peaks are present in the current trace. This approach is more robust against noise and switches than the simple peak-finding package used in Ref.~\cite{Moon2020} (see Supplementary Material). If Coulomb peaks are found by the classifier then a low-resolution current map ($16 \times 16$ pixels, 5~mV/pixel for the nanowire and 9~mV/pixel for the FinFET and the heteroturecture) is taken by sweeping the plunger gates.
    The current map is believed to contain double quantum dot features if it scores above a threshold, which is fixed and can be optimised. We use the same score function as in Ref. \cite{Moon2020}. If double quantum dot features are believed to be present, a high-resolution current map ($48 \times 48$ pixels, 4.2~mV/pixel for the nanowire and 2.5~mV/pixel for the FinFET and the heteroturecture) is taken. CATSAI runs for a certain number of iterations. A posteriori, to benchmark the algorithm's performance, humans can verify if the double quantum dot features were successfully identified by the algorithm. 
    
    After the i$\mathrm{^{th}}$ iteration, a model of the hypersurface is built using a Gaussian process, as shown in Fig. \ref{fig:fig1_devices}d--f, and $\bm{u}$ is chosen by incorporating the knowledge gained during the peak detection module in the investigation stage. The algorithm achieves this by generating a set of candidate pinch-off locations on the hypersurface and using the probability of finding Coulomb peaks in a given location of gate voltage space, $\mathrm{\tilde{P}_{peaks}}$, as a weighting for the choice of $\bm{u}$~\cite{Moon2020}. Using Thompson sampling, the algorithm then selects one of the candidate pinch-off locations, defining a new $\bf{u}$.
    In each of the following iterations, the pinch-off locations and the information gathered by the peak detection are used to update the hypersurface model and $\mathrm{\tilde{P}_{peaks}}$, respectively.
    
    CATSAI is benchmarked against a version of this algorithm which does not use a weighted hypersurface model to influence the sampling of the hypersurface. It instead continues to sample the hypersurface at random after the first i iterations, thus remaining on the left-hand branch of the sampling stage (Fig \ref{fig:fig2_algo_workflow}). We call this version of CATSAI `Random Search', although it is important to highlight that it still relies on peak detection. 
    
    \subsection*{Tuning across architectures and material systems}
    
    \begin{figure}[ht!] 
        \includegraphics[width=\columnwidth]{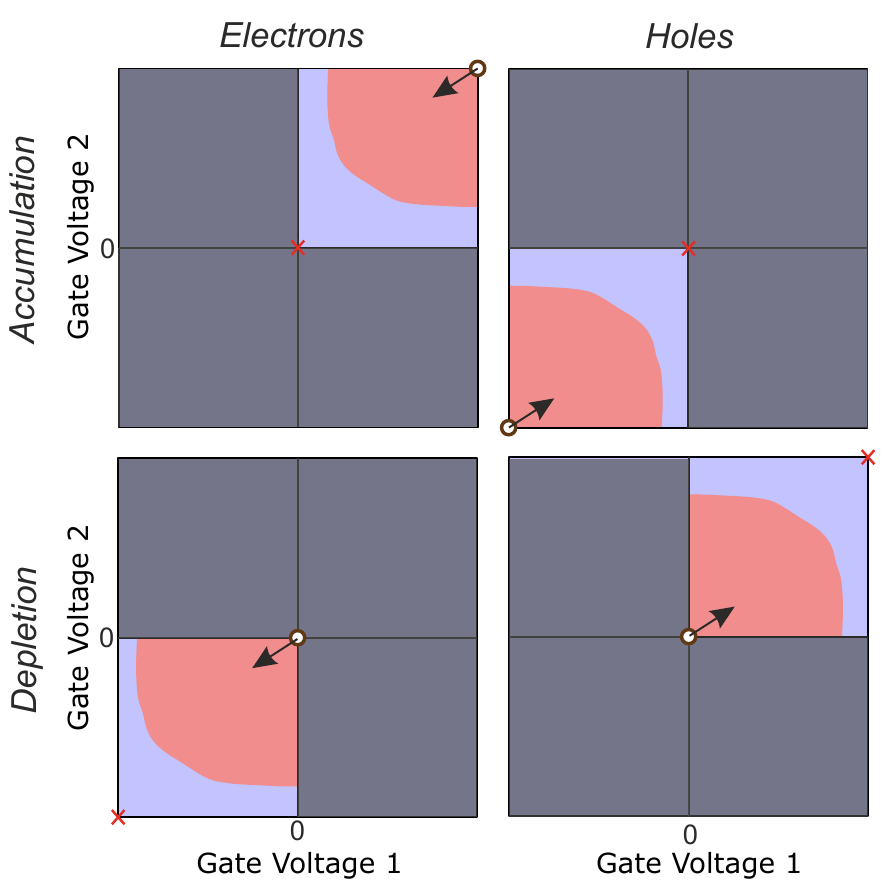}
        \caption{\label{fig:fig3_selection_rules} Gate-voltage space exploration. Different charge carriers (gate operation modes) are represented in different columns (rows). Each panel illustrates the initial placement of the origin (white circle), search boundary (red cross), and search direction (black arrow). The gate voltage space is divided into regions of near-zero (blue) and non-zero (pink) current. Regions of voltage space which cannot be explored due to the gate voltage bounds set to avoid device damage are greyed out.}
    \end{figure}
    
    To make the algorithm general across different charge carriers and modes in which gate electrodes are designed to act (depletion or accumulation), the origin, bound, and direction of the gate-voltage space exploration used in the sampling stage are set in a configuration file (Fig. \ref{fig:fig3_selection_rules}). 
    The algorithm starts in the gate voltage configuration which delivers the highest current and sweeps gate voltages in the direction of decreasing current with the aim of locating the boundary between the two regions. 
    This flexibility in the search of gate voltage space, combined with a noise-tolerant classification of Coulomb peaks in the investigation stage, makes CATSAI robust across device architectures and material systems. 
    The Coulomb peak classifier is trained on current traces acquired in different Si FinFET and GeSi nanowire devices (see Supplementary Material). This random forest classifier can successfully handle both noise and charge switches, resulting in a robust Coulomb peak detection. The number of false positives in the classification that are accepted for the next step of the investigation stage is thus reduced, significantly shortening device tuning times.

\section*{Results}

\begin{figure}[ht!] 
    \includegraphics[width=1\columnwidth]{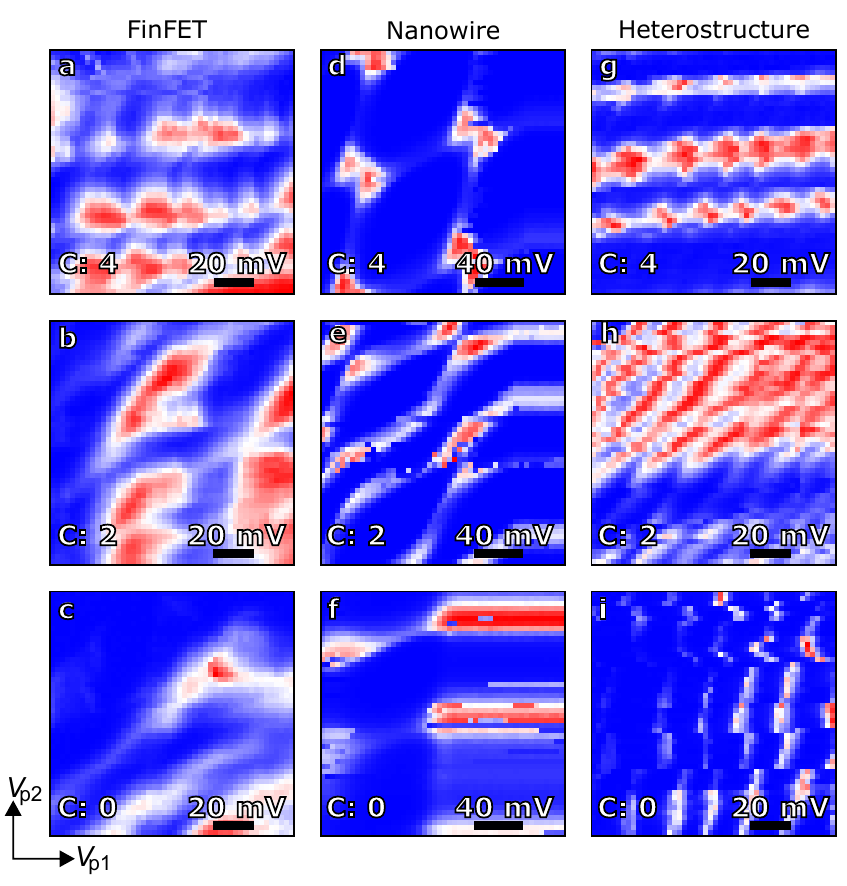}
    \caption{Device tuning. Examples of current map outputs on the different devices in which CATSAI was run. High resolution maps are generated during the investigation stage by sweeping the plunger gates of each device $V_{p1,p2}$; for the FinFET $V_{3,2}$ (a,b,c), the nanowire $V_{4,2}$ (d,e,f) and the heterostructure $V_{3,5}$ (g,h,i). These current maps are labelled \textit{a posteriori} by humans to verify whether they correspond to the double quantum dot regime. $C$ indicates the number of humans out of four who labelled the current map as corresponding to a double quantum dot regime. Red (blue) indicates regions of high (low) current in each map.}
    \label{fig:fig4_current_maps}
\end{figure}

\begin{figure*}[ht!]
    \includegraphics{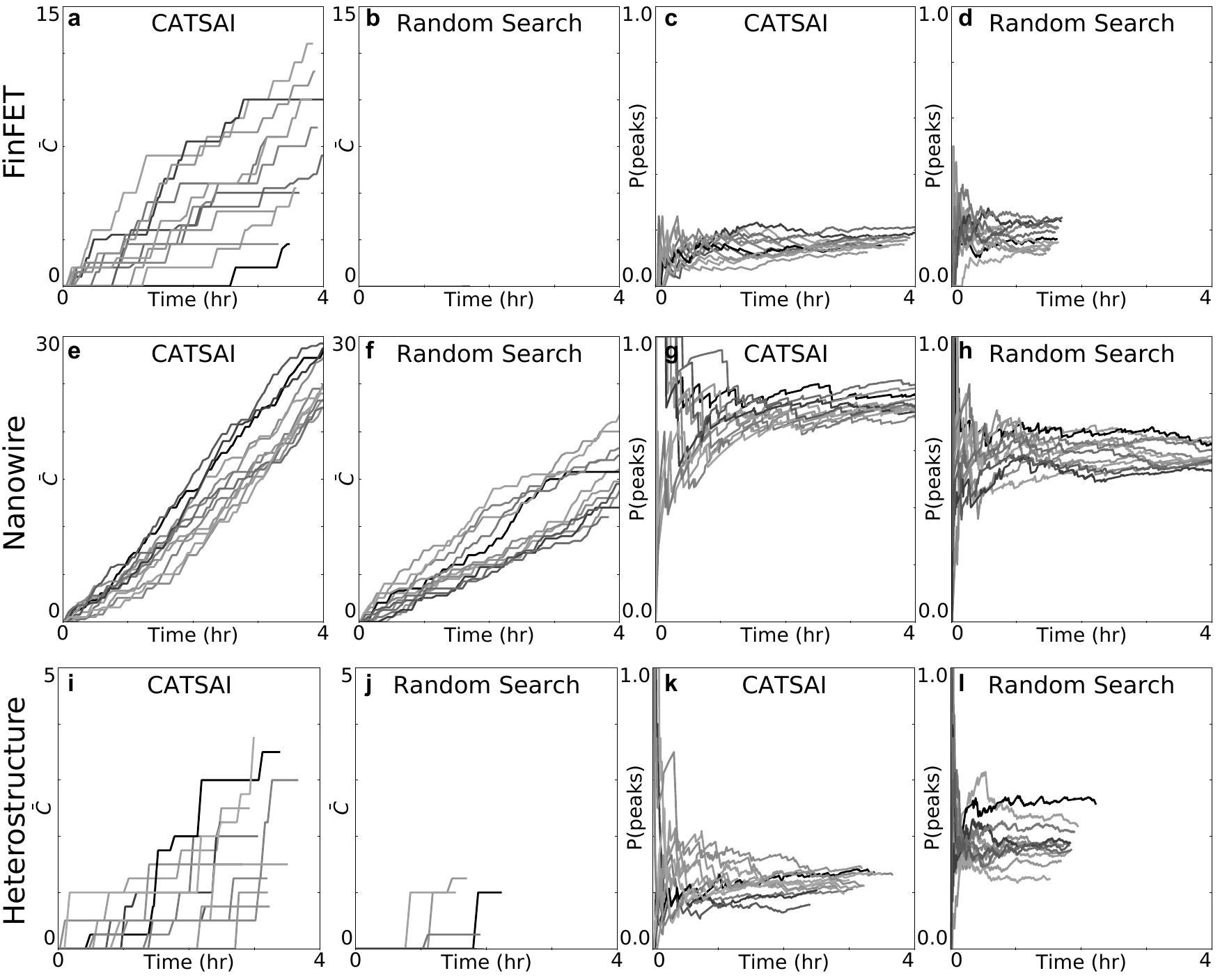}
    
    \caption{Benchmarking the algorithm's performance. Cumulative sum of the average number of double quantum dot regimes verified by humans $\bar{C}$ (first and second columns) and probability of finding Coulomb peaks  $\mathrm{P(peaks)}$ (third and fourth columns), as a function of laboratory time for each run of CATSAI and Random Search algorithms. Rows correspond to the different devices. Only the first 4 hours of each tuning run are shown for ease of visualisation. CATSAI outperforms Random Search in the number of double quantum dot regimes located for all devices. The value of $\bar{C}$ remains at 0 in many of the Random Search runs, and thus are not visible in the plots of $\bar{C}$ as a function of time. The increase in $\mathrm{P(peaks)}$ as a function of laboratory time observed for the CATSAI runs after the first 12 iterations can be explained by the algorithm `learning' a better model of the hypersurface as the Gaussian process regression acquires more observations.}
    
    \label{fig:fig5_cumulative_labels}
\end{figure*}

The algorithm was run for 250 iterations for all experiments performed. The number of iterations that the algorithm runs without an hypersurface model, i, which can be separately optimised, was fixed to twelve in this case. A few examples of output currents maps produced by CATSAI for the different devices considered are displayed in Fig.~\ref{fig:fig4_current_maps}. Although accurate most of the time, the score function that the algorithm uses to detect double quantum dot regimes can sometimes be tricked by charge switches, as observed in Fig.~\ref{fig:fig4_current_maps}i. 

To benchmark the performance of the algorithm, the output current maps were labelled by human experts at the end of the tuning experiment to verify whether they corresponded to the double quantum dot regime (see Supplementary Material). The human experts were unaware whether the current maps to be labelled were the output of CATSAI or Random Search.
We define $C$ as the number of humans who labelled a current map as containing double quantum dot features. In each iteration of the algorithm, we cumulatively sum the value of $C$ normalised by the total number of human labellers (four).  
The resulting quantity, $\bar{C}$, provides a measure of the number of double dot regimes found by the tuning algorithm while considering disagreements between human labellers.

Figure \ref{fig:fig5_cumulative_labels}a--j shows $\bar{C}$ as a function of laboratory time for 12 runs of CATSAI and Random Search for each of the devices considered. CATSAI outperforms Random Search in the total number of double quantum dot regimes located in all cases. The Random Search algorithm did relatively well in locating double quantum dot regimes in the nanowire but did not locate any double quantum dot regime in the FinFET (Fig. \ref{fig:fig5_cumulative_labels}b) and struggled to locate more than one double quantum dot regime in the SiGe heterostructure device (Fig. \ref{fig:fig5_cumulative_labels}j).

The probability of Coulomb peaks estimated for a given number of iterations, $\mathrm{P(peaks)}$, is plotted as a function of laboratory time for each algorithm run and each device in Fig. \ref{fig:fig5_cumulative_labels}c--l. For the Random Search and the first i iterations of CATSAI, the algorithm chooses pinch-off locations randomly, and thus $\mathrm{P(peaks)}$ does not show a definite trend. For the subsequent iterations, we expect CATSAI to learn which are the promising locations in gate voltage space, and $\mathrm{P(peaks)}$ should thus increase as a function of time. This increase would not be monotonic, since the algorithm balances a exploration/exploitation trade-off~\cite{Moon2020}. 

The trend of $\mathrm{P(peaks)}$ as a function of laboratory time observed in most CATSAI runs is similar for the FinFET, nanowire and the heterostructure devices. The saturation after 1--2 hours is expected given that transport feature can only be found in a limited portion of the gate voltage space. 

For the FinFET device and the heterostructure, the values of $\mathrm{P(peaks)}$ are on average larger for the Random Search than for CATSAI runs. Given we expect faster tuning times for CATSAI, either the majority of the transport features found by Random Search correspond to single quantum dots instead of double quantum dots, or the score function fails to identify double quantum dot features. The latter is unlikely to be the dominant factor given that the Random Search algorithm is run for 3000 iterations for this device and the score function is successful in identifying double dot features via CATSAI.

\begin{table}[ht]
    \begin{ruledtabular}
    \begin{tabular}{ccc}
        &\multicolumn{2}{c}{Tuning Times (minutes)}\\
        Device&CATSAI&Random Search \\ \hline
        GeSi Nanowire &9.5 (6.7, 12)&17 (9.9, 26)\\
        Si FinFET &30 (26, 37)& - \\
        SiGe Het. &92 (71, 120)&360 (190, 830) \\
    \end{tabular}
    \end{ruledtabular}
        \caption{\label{tab:tuning_times} Median device tuning times with 80\% credible intervals (equal tailed) corresponding to CATSAI and Random Search algorithm runs for all devices considered.  We estimate these credible intervals as described in Ref. \cite{Moon2020}. The Random Search tuning time for the FinFET is unknown as no double quantum dot regimes were located.} 
\end{table}
CATSAI tuned all devices faster than Random Search. The median tuning times are 10 minutes for the nanowire, 30 minutes for the FinFET, and 90 minutes for the heterostructure (Table \ref{tab:tuning_times}). The Random Search algorithm was surprisingly quick at tuning the nanowire, while unable to tune the FinFET successfully within 12 runs of the algorithm, which totals a laboratory time of 19 hours. Reduced tuning times for the FinFET device could probably be achieved by fixing the lead gate voltages. The difference between the upper and lower credible interval of the tuning times achieved in the heterostructure device is an order of magnitude less than that achieved by Random Search.

The difference between median tuning times for different devices begs the question whether the dimensionality of the gate voltage space is the key factor affecting tuning times or if there is a more subtle characteristic at play. The faster median tuning times were achieved in those devices for which the gate voltage space has the fewer dimensions, i.e. the FinFET and the nanowire. Although the nanowire does have greater gate electrode dimensionality than the FinFET, we still observe faster tuning times for the nanowire. There would seem to be more double quantum dot regimes in the nanowire gate voltage space than there are in that of the FinFET.

This hypothesis is reinforced by the lack of double quantum dot regimes found in the FinFET by Random Search and it is in agreement with the experience of human experts when tuning these devices. 

A reason for the lack of double quantum dot regimes is the sharp pinch-off that occurs as a function of the lead gate electrodes. The probability of finding lead gate voltages that enable current flow and plunger gate voltages that lead to double quantum dot regimes is inherently low. As mentioned previously, faster tuning times for FinFETS would thus be expected for CATSAI and Random Search if the lead gate voltages, $V_{1}$ and $V_{4}$, are fixed.

\section*{Conclusion}
CATSAI is the first to allow for the tuning of quantum devices across material compositions and gate architectures. We achieved tuning times faster than that of human experts in a Si FinFET, a GeSi nanowire and a SiGe heterostructure device. The tuning times reported are as low as 30, 10 and 92 minutes respectively. The capability to tune these devices from scratch completely automatically, prepares the pathway laid out for the scaling of semiconductor qubits that lend themselves to industrial scale manufacture.

An analysis of the hypersurfaces corresponding to different device types and material systems could minimise variability and boost device performance by an informed device design. The size of the gate voltage space is also an important consideration in this context. While the FinFET and the nanowire gate-voltage spaces at mv resolution have approximately $10^{14}$ and $10^{17}$ pixels respectively, the mean tuning times are only different by a factor of 3, and surprisingly the median tuning time is shorter for the nanowire device.

The heterostructure, with a gate-voltage space at mv resolution of $10^{23}$ pixels, shows a mean tuning time only 3 times longer than the nanowire. This would suggest that other factors, such as the design of the gate architecture and the disorder potential, might have a very significant role in how quickly a device can be tuned. Faster tuning times could be achieved by using device information, for example by grouping gate electrodes with similar functions. While the size of the gate voltage space is determined both by device properties and fabrication methods, the volume of the hypersurface and the volume of gate voltage space in which transport features are found could be useful to quantify device variability and to characterise and design different device architectures.

Radio-frequency reflectometry measurements would also lead to faster tuning times and the possibility of efficiently tuning large device arrays. Our work evidences the potential of machine learning-based algorithms to find overarching solutions for the control of complex quantum circuits.

\section*{Data Availability}
    The data acquired by the algorithm during experiments is available from the corresponding author upon reasonable request.
    

\begin{acknowledgments}
    We acknowledge Ang Li, Erik P. A. M. Bakkers (University of Eindhoven) for the fabrication of the Ge/Si nanowire. This work was supported by the Royal Society, the EPSRC National Quantum Technology Hub in Networked Quantum Information Technology (EP/M013243/1), Quantum Technology Capital (EP/N014995/1), EPSRC Platform Grant (EP/R029229/1), the European Research Council (Grant agreement 948932), the Swiss Nanoscience Institute, the NCCR SPIN, the EU H2020 European Microkelvin Platform EMP grant No. 824109, the Scientific Service Units of IST Austria through resources provided by the nanofabrication facility and, the FWF-P 30207 project. This publication was also made possible through support from Templeton World Charity Foundation and John Templeton Foundation. The opinions expressed in this publication are those of the authors and do not necessarily reflect the views of the Templeton Foundations.
\end{acknowledgments}

\vspace{20pt}
\section*{Author Contributions}

    B.S., D.T.L., L.C.C., F.N.M.F., S.G., G.K., D.M.Z. and the machine performed the experiments at the University of Basel and IST Austria. D.J., A.B., D.C., G.I., A.V.K., F.R.B., S.G., M.dK., M.J.C., S.S. contributed to the experiment and sample fabrication. B.S., D.T.L. developed the algorithm in collaboration with H.M., M.A.O and D.S.. B.S., D.T.L, L.C.C., N.A., F.V. and F.F. contributed to labelling and data analysis. The project was conceived by G.A.D.B. and N.A.. B.S., D.T.L. and N.A. wrote the manuscript. All authors commented and discussed the results.
    
\section*{Competing Interests}
    The authors declare no competing interests.
    
\section*{Correspondence}
    Correspondence and requests for materials
    should be addressed to Natalia Ares (email: \href{mailto:natalia.ares@materials.ox.ac.uk}{natalia.ares@materials.ox.ac.uk}).


\begin{thebibliography}{10}
\expandafter\ifx\csname url\endcsname\relax
  \def\url#1{\texttt{#1}}\fi
\expandafter\ifx\csname urlprefix\endcsname\relax\def\urlprefix{URL }\fi
\providecommand{\bibinfo}[2]{#2}
\providecommand{\eprint}[2][]{\url{#2}}

\bibitem{Loss1998}
\bibinfo{author}{Loss, D.} \& \bibinfo{author}{DiVincenzo, D.~P.}
\newblock \bibinfo{title}{Quantum computation with quantum dots}.
\newblock \emph{\bibinfo{journal}{Phys. Rev. A}} \textbf{\bibinfo{volume}{57}},
  \bibinfo{pages}{120--126} (\bibinfo{year}{1998}).

\bibitem{Veldhorst2017}
\bibinfo{author}{Veldhorst, M.}, \bibinfo{author}{Eenink, H. G.~J.},
  \bibinfo{author}{Yang, C.~H.} \& \bibinfo{author}{Dzurak, A.~S.}
\newblock \bibinfo{title}{{Silicon CMOS architecture for a spin-based quantum
  computer}}.
\newblock \emph{\bibinfo{journal}{Nature Communications}}
  \textbf{\bibinfo{volume}{8}}, \bibinfo{pages}{1766} (\bibinfo{year}{2017}).

\bibitem{vandersypen2017interfacing}
\bibinfo{author}{Vandersypen, L.} \emph{et~al.}
\newblock \bibinfo{title}{Interfacing spin qubits in quantum dots and
  donors$-$hot, dense, and coherent}.
\newblock \emph{\bibinfo{journal}{npj Quantum Information}}
  \textbf{\bibinfo{volume}{3}}, \bibinfo{pages}{34} (\bibinfo{year}{2017}).

\bibitem{Watson2018}
\bibinfo{author}{Watson, T.~F.} \emph{et~al.}
\newblock \bibinfo{title}{{A programmable two-qubit quantum processor in
  silicon}}.
\newblock \emph{\bibinfo{journal}{Nature}} \textbf{\bibinfo{volume}{555}},
  \bibinfo{pages}{633--637} (\bibinfo{year}{2018}).

\bibitem{Hendrickx2020}
\bibinfo{author}{Hendrickx, N.~W.} \emph{et~al.}
\newblock \bibinfo{title}{{A four-qubit germanium quantum processor}}.
\newblock \emph{\bibinfo{journal}{Nature}} \textbf{\bibinfo{volume}{591}},
  \bibinfo{pages}{580--585} (\bibinfo{year}{2021}).

\bibitem{Scappucci2020a}
\bibinfo{author}{Scappucci, G.} \emph{et~al.}
\newblock \bibinfo{title}{{The germanium quantum information route}}.
\newblock \emph{\bibinfo{journal}{Nature Reviews Materials}}
  \bibinfo{pages}{1--18} (\bibinfo{year}{2020}).

\bibitem{Takeda2021}
\bibinfo{author}{Takeda, K.} \emph{et~al.}
\newblock \bibinfo{title}{{Quantum tomography of an entangled three-qubit state
  in silicon}}.
\newblock \emph{\bibinfo{journal}{Nature Nanotechnology}} \bibinfo{pages}{1--5}
  (\bibinfo{year}{2021}).

\bibitem{Fowler2012}
\bibinfo{author}{Fowler, A.~G.}, \bibinfo{author}{Mariantoni, M.},
  \bibinfo{author}{Martinis, J.~M.} \& \bibinfo{author}{Cleland, A.~N.}
\newblock \bibinfo{title}{{Surface codes: Towards practical large-scale quantum
  computation}}.
\newblock \emph{\bibinfo{journal}{Physical Review A - Atomic, Molecular, and
  Optical Physics}} \textbf{\bibinfo{volume}{86}}, \bibinfo{pages}{032324}
  (\bibinfo{year}{2012}).

\bibitem{Preskill2018}
\bibinfo{author}{Preskill, J.}
\newblock \bibinfo{title}{{Quantum Computing in the NISQ era and beyond}}.
\newblock \emph{\bibinfo{journal}{Quantum}} \textbf{\bibinfo{volume}{2}},
  \bibinfo{pages}{79} (\bibinfo{year}{2018}).

\bibitem{Ares2021}
\bibinfo{author}{Ares, N.}
\newblock \bibinfo{title}{{Machine learning as an enabler of qubit
  scalability}}.
\newblock \emph{\bibinfo{journal}{Nature Reviews Materials}}
  \bibinfo{pages}{1--2} (\bibinfo{year}{2021}).

\bibitem{Baart2016}
\bibinfo{author}{Baart, T.~A.}, \bibinfo{author}{Eendebak, P.~T.},
  \bibinfo{author}{Reichl, C.}, \bibinfo{author}{Wegscheider, W.} \&
  \bibinfo{author}{Vandersypen, L.~M.}
\newblock \bibinfo{title}{{Computer-automated tuning of semiconductor double
  quantum dots into the single-electron regime}}.
\newblock \emph{\bibinfo{journal}{Applied Physics Letters}}
  \textbf{\bibinfo{volume}{108}}, \bibinfo{pages}{213104}
  (\bibinfo{year}{2016}).

\bibitem{Botzem2018}
\bibinfo{author}{Botzem, T.} \emph{et~al.}
\newblock \bibinfo{title}{{Tuning Methods for Semiconductor Spin Qubits}}.
\newblock \emph{\bibinfo{journal}{Physical Review Applied}}
  \textbf{\bibinfo{volume}{10}}, \bibinfo{pages}{054026}
  (\bibinfo{year}{2018}).

\bibitem{VanDiepen2018}
\bibinfo{author}{van Diepen, C.~J.} \emph{et~al.}
\newblock \bibinfo{title}{{Automated tuning of inter-dot tunnel coupling in
  double quantum dots}}.
\newblock \emph{\bibinfo{journal}{Applied Physics Letters}}
  \textbf{\bibinfo{volume}{113}}, \bibinfo{pages}{033101}
  (\bibinfo{year}{2018}).

\bibitem{Teske2019}
\bibinfo{author}{Teske, J.~D.} \emph{et~al.}
\newblock \bibinfo{title}{{A machine learning approach for automated
  fine-tuning of semiconductor spin qubits}}.
\newblock \emph{\bibinfo{journal}{Applied Physics Letters}}
  \textbf{\bibinfo{volume}{114}}, \bibinfo{pages}{133102}
  (\bibinfo{year}{2019}).

\bibitem{Volk2019}
\bibinfo{author}{Volk, C.} \emph{et~al.}
\newblock \bibinfo{title}{{Loading a quantum-dot based “Qubyte” register}}.
\newblock \emph{\bibinfo{journal}{npj Quantum Information}}
  \textbf{\bibinfo{volume}{5}}, \bibinfo{pages}{29} (\bibinfo{year}{2019}).

\bibitem{Mills2019}
\bibinfo{author}{Mills, A.~R.} \emph{et~al.}
\newblock \bibinfo{title}{{Computer-automated tuning procedures for
  semiconductor quantum dot arrays}}.
\newblock \emph{\bibinfo{journal}{Applied Physics Letters}}
  \textbf{\bibinfo{volume}{115}}, \bibinfo{pages}{113501}
  (\bibinfo{year}{2019}).

\bibitem{Kalantre2019}
\bibinfo{author}{Kalantre, S.~S.} \emph{et~al.}
\newblock \bibinfo{title}{{Machine Learning techniques for state recognition
  and auto-tuning in quantum dots}}.
\newblock \emph{\bibinfo{journal}{npj Quantum Information}}
  \textbf{\bibinfo{volume}{5}}, \bibinfo{pages}{6} (\bibinfo{year}{2019}).

\bibitem{Zwolak2020}
\bibinfo{author}{Zwolak, J.~P.} \emph{et~al.}
\newblock \bibinfo{title}{{Autotuning of Double-Dot Devices In Situ with
  Machine Learning}}.
\newblock \emph{\bibinfo{journal}{Physical Review Applied}}
  \textbf{\bibinfo{volume}{13}}, \bibinfo{pages}{034075}
  (\bibinfo{year}{2020}).

\bibitem{Durrer2020}
\bibinfo{author}{Durrer, R.} \emph{et~al.}
\newblock \bibinfo{title}{{Automated Tuning of Double Quantum Dots into
  Specific Charge States Using Neural Networks}}.
\newblock \emph{\bibinfo{journal}{Physical Review Applied}}
  \textbf{\bibinfo{volume}{13}}, \bibinfo{pages}{054019}
  (\bibinfo{year}{2020}).

\bibitem{Darulova2019}
\bibinfo{author}{Darulov{\'{a}}, J.} \emph{et~al.}
\newblock \bibinfo{title}{{Autonomous tuning and charge state detection of gate
  defined quantum dots}}.
\newblock \emph{\bibinfo{journal}{Physical Review Applied}}
  \textbf{\bibinfo{volume}{13}}, \bibinfo{pages}{054005}
  (\bibinfo{year}{2019}).

\bibitem{Moon2020}
\bibinfo{author}{Moon, H.} \emph{et~al.}
\newblock \bibinfo{title}{{Machine learning enables completely automatic tuning
  of a quantum device faster than human experts}}.
\newblock \emph{\bibinfo{journal}{Nature Communications}}
  \textbf{\bibinfo{volume}{11}}, \bibinfo{pages}{4161} (\bibinfo{year}{2020}).

\bibitem{Kuhlmann2018}
\bibinfo{author}{Kuhlmann, A.~V.}, \bibinfo{author}{Deshpande, V.},
  \bibinfo{author}{Camenzind, L.~C.}, \bibinfo{author}{Zumb{\"{u}}hl, D.~M.} \&
  \bibinfo{author}{Fuhrer, A.}
\newblock \bibinfo{title}{{Ambipolar quantum dots in undoped silicon fin
  field-effect transistors}}.
\newblock \emph{\bibinfo{journal}{Applied Physics Letters}}
  \textbf{\bibinfo{volume}{113}}, \bibinfo{pages}{122107}
  (\bibinfo{year}{2018}).

\bibitem{Camenzind2021a}
\bibinfo{author}{Camenzind, L.~C.} \emph{et~al.}
\newblock \bibinfo{title}{{A spin qubit in a fin field-effect transistor.}}
  \bibinfo{pages}{Preprint at \url{http://arxiv.org/abs/2103.07369}}
  (\bibinfo{year}{2021}).

\bibitem{Geyer2021}
\bibinfo{author}{Geyer, S.} \emph{et~al.}
\newblock \bibinfo{title}{{Self-aligned gates for scalable silicon quantum
  computing}}.
\newblock \emph{\bibinfo{journal}{Applied Physics Letters}}
  \textbf{\bibinfo{volume}{118}}, \bibinfo{pages}{104004}
  (\bibinfo{year}{2021}).

\bibitem{Froning2018}
\bibinfo{author}{Froning, F.~N.} \emph{et~al.}
\newblock \bibinfo{title}{{Single, double, and triple quantum dots in Ge/Si
  nanowires}}.
\newblock \emph{\bibinfo{journal}{Applied Physics Letters}}
  \textbf{\bibinfo{volume}{113}}, \bibinfo{pages}{73102}
  (\bibinfo{year}{2018}).

\bibitem{Froning2021a}
\bibinfo{author}{Froning, F.~N.} \emph{et~al.}
\newblock \bibinfo{title}{{Ultrafast hole spin qubit with gate-tunable
  spin–orbit switch functionality}}.
\newblock \emph{\bibinfo{journal}{Nature Nanotechnology}}
  \textbf{\bibinfo{volume}{16}}, \bibinfo{pages}{308--312}
  (\bibinfo{year}{2021}).

\bibitem{Froning2021}
\bibinfo{author}{Froning, F. N.~M.} \emph{et~al.}
\newblock \bibinfo{title}{{ Strong spin-orbit interaction and g -factor
  renormalization of hole spins in Ge/Si nanowire quantum dots }}.
\newblock \emph{\bibinfo{journal}{Physical Review Research}}
  \textbf{\bibinfo{volume}{3}}, \bibinfo{pages}{13081} (\bibinfo{year}{2021}).

\bibitem{Kuhlmann2018a}
\bibinfo{author}{Kuhlmann, A.~V.}, \bibinfo{author}{Deshpande, V.},
  \bibinfo{author}{Camenzind, L.~C.}, \bibinfo{author}{Zumb{\"{u}}hl, D.~M.} \&
  \bibinfo{author}{Fuhrer, A.}
\newblock \bibinfo{title}{{Ambipolar quantum dots in undoped silicon fin
  field-effect transistors}}.
\newblock \emph{\bibinfo{journal}{Applied Physics Letters}}
  \textbf{\bibinfo{volume}{113}}, \bibinfo{pages}{122107}
  (\bibinfo{year}{2018}).

\bibitem{Breiman2001}
\bibinfo{author}{Breiman, L.}
\newblock \bibinfo{title}{{Random forests}}.
\newblock \emph{\bibinfo{journal}{Machine Learning}}
  \textbf{\bibinfo{volume}{45}}, \bibinfo{pages}{5--32} (\bibinfo{year}{2001}).

\bibitem{scikit-learn}
\bibinfo{author}{Pedregosa, F.} \emph{et~al.}
\newblock \bibinfo{title}{Scikit-learn: Machine learning in {P}ython}.
\newblock \emph{\bibinfo{journal}{Journal of Machine Learning Research}}
  \textbf{\bibinfo{volume}{12}}, \bibinfo{pages}{2825--2830}
  (\bibinfo{year}{2011}).

\end{thebibliography}

    
\clearpage

\section*{Supplementary Material}
\setcounter{figure}{0}    
\setcounter{table}{0}
\renewcommand{\figurename}{Supplementary Figure}
\renewcommand{\tablename}{Supplementary Table}

\renewcommand\tablename{Supplementary Table}
\renewcommand\figurename{Supplementary Figure}

\subsection*{Supplementary Methods}
    \subsubsection*{3D hypersurface plots}
    The 3D plot of the hypersurface for each device was generated by relying on the same method that CATSAI uses to generate the hypersurface of each device as it proceeds to coarsely tune it. The main difference being that no sampling is involved; the surface is generated by a model that makes use of the pinch-off locations detected during an algorithm run selected at random (CATSAI run 10). The model of the hypersurface used was a Gaussian Process (Matern52 Kernel). This model is then used as an interpolation method to generate the 3D plots; regularly spaced points in gate voltage space are considered and the model is used to determine whether these points lie within the hypersurface. The gate voltages not considered for the plots are kept constant at their respective mean gate voltage values for which pinch-off was observed during the experiment (Supplementary Table \ref{tab:hypersuface_bounds}).
    
    \subsubsection*{Coulomb peak detector}

    Due to the different types of current noise observed for each of the devices considered, a robust Coulomb peak detector was required. We thus developed a random forest Coulomb peak classifier.
    
    A set of 128-pixel current traces was obtained running the tuning algorithm developed by Moon et al. \cite{Moon2020} on different devices to those for which CATSAI was tested (Supplementary Table \ref{tab:test_train_table}); two different 5-gate GeSi nanowires (400~mV-long current traces), and a single 3-gate Si FinFET (200~mV-long current traces). We gathered 1095 current traces from GeSi nanowire device 1, 1321 from GeSi nanowire device 2, and 4306 from the Si FinFET device 1. The 6722 current traces were labelled by a single labeller (Brandon Severin), from which there were 553 labelled as positive (current traces containing Coulomb peaks) and the remainder (6169 current traces) were labelled as negative. 553 negative examples were randomly picked from the shuffled 6169 negative examples, to make up an even dataset of 1106 current traces. The breakdown of the data subsets include, for the positives: 115 traces from GeSi nanowire device 1, 100 from GeSi nanowire device 2 and 338 from the Si FinFET device 1. For the negative subset: 83 from GeSi nanowire device 1, 113 from GeSi nanowire device 2, and 357 from the Si FinFET device 1. Randomly chosen current traces from the even dataset of 1106 current traces were used to train and test the random forest Coulomb peak classifier; 70\% of the traces chosen were used to train the classifier, and 30\% were used to test it. No characteristic feature engineering or data pre-processing was done other than normalisation. The characteristic features the random forest classifier was trained on were the normalised current values of each trace at each pixel point, thus each sample had 128 characteristic features. The classifier relies on the Scikit-learn's ensemble RandomForestClassifier package \cite{scikit-learn}. An accuracy of 84\% was achieved. The random forest classifier was then retested on 1562 current traces from a 5-gate GeSi heterostructure device 1 and an accuracy of 92\% was achieved (Supplementary Table \ref{tab:test_train_table}, Test 2). This relatively high accuracy contrasts the Coulomb peak detector used in Ref.~\cite{Moon2020}, which achieved an accuracy of 20\% classifying the current traces obtained for the GeSi heterostructure device 1.
    
    \subsubsection*{Algorithm configuration for the different type of devices studied}
    
   Across all devices the initialisation of the algorithm is set to 12 iterations (the first 5\% of the total number of iterations for each run). In this work we did not apply any pruning rules~\cite{Moon2020}. When searching for the hypersurface, the algorithm looks for current drops below $0.5\%$ of the maximum current range. The parameters chosen to run the algorithm can be separately optimised. The model of the hypersurface is built via a Gaussian Process as in Ref.~\cite{Moon2020}.
    
    Other configuration parameters depend on the type of device to be explored (Supplementary Table \ref{tab:tuning_bounds} \& \ref{tab:config_differences}). These parameters include: voltage bounds (origin and limit) set for each gate electrode to prevent device damage, the value at which the bias voltage is fixed, the noise and segmentation thresholds, and the size in gate voltage space of current traces (diag\_trace), as well as low and high resolution current maps (2d\_lowres and 2d\_highres, respectively).
    
   During the investigation stage the current traces have a length of 128 pixels, the low resolution current maps have a size of 16 $\times$ 16 pixels, and the high resolution have a size of 48 $\times$ 48 pixels. The dimensions of the traces and the scans in voltage space are device dependent (Supplementary Table \ref{tab:config_differences}).
    
    The bias voltages were chosen to be slightly larger than typical charging energies expected for single quantum dots in each device. The noise and segmentation thresholds were chosen according to expected values; these can easily be replaced by a fixed percentage of the maximum-minimum current range across devices. The size of current traces and current maps in the investigation stage was larger for the GeSi nanowires, since the gate lever arms in these devices is often smaller compared to the other devices. These hyperparameters could also be optimised in future implementations.
    \subsubsection*{Labelling procedure}
    The current maps that are classified by the Algorithm as corresponding to a double dot regime are checked and labelled by human beings at the end of the experiment to benchmark the Algorithm's performance (Supplementary Table \ref{tab:labelling_details_FD} \& \ref{tab:labelling_details_PR}). There is often disagreement between humans about what current maps correspond to a double quantum dot regime. The current maps for each type of device were thus labelled by 4 different and independent human labellers. Three datasets were collected, one for each device (nanowire, heterostructure and FinFET). For each device, the current maps collected by Random Search and CATSAI were grouped together and shuffled to avoid labellers' confirmation bias. Median tuning times were calculated using a Bayesian model based on the resultant labels as in Ref.~\cite{Moon2020}.






\subsection*{Supplementary Tables}

    \begin{table}[ht]
    \begin{ruledtabular}
        \begin{tabular}{ccccc}
        Device               & Train & Test 1 & Test 2 & Algorithm run \\ \hline
        GeSi Nanowire 0      & -     & -      & -      &x\\ 
        GeSi Nanowire 1      & x     & x      & -      &-\\ 
        GeSi Nanowire 2      & x     & x      & -      &-\\ 
        Si FinFET 0          & -     &-      & -    & x \\ 
        Si FinFET 1          & x     & x      & -     & - \\ 
        GeSi Heterostructure  0 & -     & -      & -   & x  \\
        GeSi Heterostructure  1 & -     & -      & x   & -   
        \end{tabular}
    \end{ruledtabular}
     \caption{\label{tab:test_train_table} Devices used throughout this work. All devices used for training and or testing are different to the devices used in the experiment. Devices used for the experiment algorithm runs only are numbered as zero.}
\end{table}

    \begin{table*}[ht!]
        \caption{\label{tab:hypersuface_bounds} Bounds used for the 3D hypersurface plots.} 
        \begin{ruledtabular}
        \begin{tabular}{cccccccc}
            Device&$V_1$ (V)&$V_2$ (V)&$V_3$ (V)&$V_4$ (V)&$V_5$ (V)&$V_6$ (V)&$V_7$ \\ \hline
    Si FinFET, origin & -6.5 & -1.5 & -1.5 & -5.0 & - & - & - \\
    Si FinFET, limit & -2.5 & 0.0 & 0.0 & -5.0& - & - & - \\
    \hline
    GeSi Nanowire, origin & 0.0 & 0.56 & 0.0 & 1.1 & 0.0 & - & - \\
    GeSi Nanowire, limit & 4.0 & 0.56 & 2.5 & 1.1 & 4.0 & - & -\\
    \hline
    SiGe Heterostructure, origin & 0.48 & 0.0 & 0.74 & 0.0 & 0.79 & 0.0 & 0.41 \\
    SiGe Heterostructure, limit & 0.48 & 2.0 & 0.74 & 2.0 & 0.79 & 2.0 & 0.41 \\
    
        \end{tabular}
        \end{ruledtabular}
    \end{table*}
    
    \begin{table*}[ht!]
        \caption{\label{tab:tuning_bounds} Gate voltage space explored by CATSAI and Random Search algorithms for each of the devices considered.} 
        \begin{ruledtabular}
        \begin{tabular}{cccccccc}
            Device&$V_1$ (V)&$V_2$ (V)&$V_3$ (V)&$V_4$ (V)&$V_5$ (V)&$V_6$ (V)&$V_7$ (V)\\ \hline
    Si FinFET, origin & -6.5 & -1.5 & -1.5 & -6.5 & - & - & -\\
    Si FinFET, limit & 0.0 & 0.0 & 0.0 & 0.0 & - & - & - \\
    \hline
    GeSi Nanowire, origin & 0.0 & 0.0 & 0.0 & 0.0 & 0.0 & - & -\\
    GeSi Nanowire, limit & 4.0 & 2.5 & 2.5 & 4.0& 4.0 & - & -\\
    \hline
    SiGe Heterostructure, origin & 0.0 & 0.0 & 0.0 & 0.0 & 0.0 & 0.0 & 0.0\\
    SiGe Heterostructure, limit & 2.0 & 2.0 & 2.0 & 2.0 & 2.0 & 2.0 & 2.0\\
    
        \end{tabular}
        \end{ruledtabular}
    \end{table*}

    \begin{table*}[ht!]
        \caption{\label{tab:config_differences}Differences in the configuration of the algorithm for each of the devices considered.} %
        
        \begin{ruledtabular}
            \begin{tabular}{ccccccc}
                Device &
                  $V_{bias}$ (mV) &
                  \begin{tabular}[c]{@{}c@{}}Noise \\ Threshold (pA)\end{tabular} &
                  \begin{tabular}[c]{@{}c@{}}Segmentation \\ Threshold (pA)\end{tabular} &
                  \begin{tabular}[c]{@{}c@{}}diag\_trace: \\ size (mV)\end{tabular} &
                  \begin{tabular}[c]{@{}c@{}}2d\_lowres: \\ size (mV)\end{tabular} &
                  \begin{tabular}[c]{@{}c@{}}2d\_highres: \\ size (mV)\end{tabular} \\
                  \hline
                Si FinFET            & 
                7.6   & 2 & 
                20 & 100 &
                80 $\times$ 80 &
                120 $\times$ 120 \\
                GeSi Nanowire        & 4   
                & 2 & 1000
                & 200  & 150 $\times$ 150 &
                200 $\times$ 200 \\
                SiGe Heterostructure & 0.5 & 
                10 & 30
                & 100 & 80 $\times$ 80 & 120 $\times$ 120
            \end{tabular}
        \end{ruledtabular}
    \end{table*}

    \begin{table*}[ht!]
        \caption{\label{tab:labelling_details_FD} Total number of current maps labelled as positive (i.e. corresponding to the double quantum dot regime) found by each labeller (Labeller 1, 2, 3, 4) for each device and for each run of CATSAI. Runs marked with a an asterisk were excluded because the cryostat temperature was slightly higher than base temperature. 
        } 
        \begin{ruledtabular}
        \begin{tabular}{ccccccc}
            Experiment&Iterations&Time (hours)& Labeller 1 & Labeller 2 & Labeller 3 & Labeller 4\\ \hline
    Si FinFET, run 1 & 250 & 3.47 & 2 & 2 & 2 & 3\\
    Si FinFET, run 2 & 250 & 4.17 & 12 & 12 & 10 & 10\\
    Si FinFET, run 3 & 250 & 3.62 & 5 & 5 & 5 & 5\\
    Si FinFET, run 4 & 250 & 4.15 & 9 & 6 & 6 & 7\\
    Si FinFET, run 5 & 250 & 3.30 & 9 & 9 & 6 & 8\\
    Si FinFET, run 6 & 250 & 3.90 & 9 & 9 & 7 & 9\\
    Si FinFET, run 7 & 250 & 3.30 & 3 & 2 & 1 & 3\\
    Si FinFET, run 8 & 250 & 3.86 & 13 & 13 & 7 & 13\\
    Si FinFET, run 9 & 250 & 3.25 & 4 & 4 & 4 & 4\\
    Si FinFET, run 10 & 250 & 3.81 & 10 & 11 & 8 & 11\\
    Si FinFET, run 11 & 250 & 3.57 & 5 & 5 & 5 & 6\\
    Si FinFET, run 12 & 250 & 3.83 & 13 & 13 & 13 & 13\\
    \hline
    GeSi Nanowire, run 1 & 250 & 8.42 & 45 & 58 & 74 & 48\\
    GeSi Nanowire, run 2 & 250 & 8.26 & 46 & 61 & 80 & 54\\
    GeSi Nanowire, run 3 & 250 & 8.57 & 38 & 60 & 77 & 49\\
    GeSi Nanowire, run 4 & 250 & 9.18 & 40 & 64 & 79 & 46\\
    GeSi Nanowire, run 5 & 250 & 8.21 & 38 & 52 & 73 & 47\\
    GeSi Nanowire, run 6 & 250 & 8.90 & 38 & 64 & 78 & 54\\
    GeSi Nanowire, run 7 & 250 & 8.12 & 39 & 46 & 70 & 46\\
    GeSi Nanowire, run 8 & 250 & 8.68 & 46 & 59 & 79 & 48\\
    GeSi Nanowire, run 9 & 250 & 9.05 & 50 & 67 & 84 & 48\\
    GeSi Nanowire, run 10 & 250 & 9.31 & 51 & 64 & 78 & 52\\
    GeSi Nanowire, run 11 & 250 & 9.38 & 50 & 64 & 82 & 54\\
    GeSi Nanowire, run 12 & 250 & 9.02 & 43 & 63 & 78 & 55\\
    \hline
    SiGe Heterostructure, run 1 & 250 & 3.38 & 2 & 4 & 5 & 3\\
    SiGe Heterostructure, run 2 & 250 & 2.50 & 2 & 3 & 2 & 2\\
    SiGe Heterostructure, run 3 & 250 & 2.39 & 1 & 1 & 0 & 1\\
    SiGe Heterostructure, run 4* & 250 & 3.17 & 1 & 2 & 0 & 1\\
    SiGe Heterostructure, run 5 & 250 & 3.04 & 3 & 2 & 2 & 1\\
    SiGe Heterostructure, run 6 & 250 & 3.66 & 2 & 3 & 4 & 3\\
    SiGe Heterostructure, run 7 & 250 & 3.19 & 1 & 1 & 1 & 2\\
    SiGe Heterostructure, run 8 & 250 & 2.81 & 2 & 1 & 2 & 1\\
    SiGe Heterostructure, run 9 & 250 & 3.19 & 1 & 1 & 1 & 1\\
    SiGe Heterostructure, run 10 & 250 &3.22 & 1 & 0 & 1 & 1\\
    SiGe Heterostructure, run 11 & 250 &2.91 & 3 & 4 & 1 & 2\\
    SiGe Heterostructure, run 12 & 250 &3.50 & 1 & 2 & 2 & 1\\
    SiGe Heterostructure, run 13* & 250 &3.42 & 2 & 2 & 2 & 3\\
    SiGe Heterostructure, run 14* & 250 &3.31 & 4 & 3 & 5 & 3\\
    SiGe Heterostructure, run 15 & 250 &2.99 & 3 & 4 & 4 & 4\\
    
        \end{tabular}
        \end{ruledtabular}
    \end{table*}
    
    \begin{table*}[ht!]
        \caption{\label{tab:labelling_details_PR} Total number of current maps labelled as positive (i.e. corresponding to the double quantum dot regime) found by each labeller (Labeller 1, 2, 3, 4) for each device and for each run of Random Search. Runs marked with a an asterisk were excluded because the cryostat temperature was slightly higher than base temperature. 
        } 
        \begin{ruledtabular}
        \begin{tabular}{ccccccc}
            Experiment&Iterations&Time (hours)& Labeller 1 & Labeller 2 & Labeller 3 & Labeller 4\\ \hline
    Si FinFET, run 1 & 250 & 1.62 &  0 & 0 & 0 & 0\\
    Si FinFET, run 2 & 250 & 1.68 &  0 & 0 & 0 & 0\\
    Si FinFET, run 3 & 250 & 1.69 &  0 & 0 & 0 & 0\\
    Si FinFET, run 4 & 250 & 1.58 &  0 & 0 & 0 & 0\\
    Si FinFET, run 5 & 250 & 1.64 &  0 & 0 & 0 & 0\\
    Si FinFET, run 6 & 250 & 1.62 &  0 & 0 & 0 & 0\\
    Si FinFET, run 7 & 250 & 1.51 &  0 & 0 & 0 & 0\\
    Si FinFET, run 8 & 250 & 1.45 &  0 & 0 & 0 & 0\\
    Si FinFET, run 9 & 250 & 1.49 &  0 & 0 & 0 & 0\\
    Si FinFET, run 10 & 250 &1.52 & 0 & 0 & 0 & 0\\
    Si FinFET, run 11 & 250 &1.63 & 0 & 0 & 0 & 0\\
    Si FinFET, run 12 & 250 &1.56 & 0 & 0 & 0 & 0\\
    \hline
    GeSi Nanowire, run 1 & 250 & 4.40 & 11 & 18 & 23 & 15\\
    GeSi Nanowire, run 2 & 250 & 4.06 & 5  & 13 & 20 & 10\\
    GeSi Nanowire, run 3 & 250 & 4.44 & 9  & 17 & 28 & 11\\
    GeSi Nanowire, run 4 & 250 & 3.82 & 3  & 12 & 21 & 8\\
    GeSi Nanowire, run 5 & 250 & 4.66 & 12 & 20 & 30 & 14\\
    GeSi Nanowire, run 6 & 250 & 4.58 & 10 & 22 & 32 & 17\\
    GeSi Nanowire, run 7 & 250 & 4.17 & 11 & 11 & 22 & 13\\
    GeSi Nanowire, run 8 & 250 & 3.92 & 7  & 14 & 21 & 10\\
    GeSi Nanowire, run 9 & 250 & 4.53 & 14 & 23 & 30 & 17\\
    GeSi Nanowire, run 10 & 250 &4.37 & 12 & 19 & 23 & 16\\
    GeSi Nanowire, run 11 & 250 &4.59 & 11 & 20 & 30 & 14\\
    GeSi Nanowire, run 12 & 250 &4.21 & 19 & 23 & 28 & 18\\
    \hline
    SiGe Heterostructure, run 1 & 250 & 2.22 & 1 & 1 & 1 & 1\\
    SiGe Heterostructure, run 2 & 250 & 1.83 & 0 & 0 & 0 & 0\\
    SiGe Heterostructure, run 3 & 250 & 1.82 & 0 & 0 & 0 & 0\\
    SiGe Heterostructure, run 4 & 250 & 1.85 & 0 & 0 & 0 & 0\\
    SiGe Heterostructure, run 5 & 250 & 1.89 & 0 & 1 & 0 & 0\\
    SiGe Heterostructure, run 6 & 250 & 1.82 & 0 & 0 & 0 & 0\\
    SiGe Heterostructure, run 7 & 250 & 1.72 & 0 & 0 & 0 & 0\\
    SiGe Heterostructure, run 8 & 250 & 1.68 & 0 & 0 & 0 & 0\\
    SiGe Heterostructure, run 9 & 250 & 1.69 & 1 & 2 & 1 & 1\\
    SiGe Heterostructure, run 10 & 250 &1.81 & 0 & 0 & 0 & 0\\
    SiGe Heterostructure, run 11 & 250 &1.95 & 0 & 0 & 0 & 0\\
    SiGe Heterostructure, run 12 & 250 &1.52 & 1 & 1 & 1 & 1\\
    SiGe Heterostructure, run 13* & 250 &1.64 & 0 & 0 & 1 & 0\\

        \end{tabular}
        \end{ruledtabular}
    \end{table*}

\end{document}